\documentstyle[12pt,epsf]{article}
\def\beq{\begin{equation}}   \def\eeq{\end{equation}}

\def\a{{\rm AdS_5}}

\begin{document}
\thispagestyle{empty}
\begin{flushright}
NYU-TH/99/10/02 \\
\end{flushright}

\vspace{0.1in}
\begin{center}
\bigskip\bigskip
{\Large \bf Localization of Matter and  Cosmological Constant 
on a Brane in Anti de Sitter Space}

\vspace{0.3in}

{Borut Bajc\footnote{On leave of absence from 
J. Stefan Institute, Ljubljana, Slovenia} 
and Gregory Gabadadze}
\vspace{0.1in}

{\baselineskip=14pt \it 
Department of Physics, New York University, New York, NY 10003 } \\
\vspace{0.2in}
\end{center}

\vspace{0.9cm}
\begin{center}
{\bf Abstract}
\end{center} 
\vspace{0.1in}

We study two issues, the localization of 
various spin fields, and the problem of the cosmological constant on a
brane in five-dimensional anti de Sitter space.  
We find  that spin-zero fields are localized on a positive-tension brane.
In addition to the localized zero-mode  there is a continuous tower
of states with no mass gap. 
Spin one-half and three-half states can be  localized on a brane with 
``negative tension''. Their localization
can be achieved on the  positive-tension brane as well, if  
additional interactions are introduced. 
The necessary ingredient 
of the scenario  with localized gravity
is the relation between the bulk cosmological constant and
the brane tension. In the absence of supersymmetry  
this implies fine-tuning between the parameters of the theory.
To deal with this issue we introduce
a four-form gauge field. This  gives 
an additional arbitrary contribution to the bulk 
cosmological constant. As a result, the model  gives rise to  a  
continuous family of brane Universe solutions 
for generic values of the bulk cosmological constant and the 
brane tension.
Among these solutions there  is one with a zero four-dimensional 
cosmological constant.

\newpage

\section{Introduction and results}

Large extra dimensions offer an opportunity for a new solution
to  the hierarchy problem \cite {ADD}. 
The crucial  ingredient of this scenario is a brane on which standard
model particles are localized. 
Field-theoretic localization mechanisms for  
scalars and fermions \cite {RubakovShaposhnikov} as well as 
for gauge bosons \cite {DvaliShifman} were found.
In string theory, fields 
can naturally be  localized on D-branes  
due to the open strings ending on them \cite {Polchinski}
(for string theory realization of the scenario of Ref. \cite {ADD},
see, e.g., \cite {Antoni,KTye}).  

Up until recently, extra dimensions had to be  compactified,
since the localization mechanism for gravity was not known.  
It was suggested in Ref. \cite {Gogber} 
that gravitational interactions
between particles on a brane in uncompactified five-dimensional space 
could  have  the correct four-dimensional Newtonian behavior, 
provided that the bulk cosmological constant and the 
brane tension are related. 
Recently, it was found by Randall and Sundrum that 
gravitons can be  localized on a brane 
which separates two patches  of $\a$ space-time \cite {RandallSundrum}. 
The necessary requirement for the four-dimensional 
brane Universe to be static is 
that  the tension of the brane
is fine-tuned to the bulk cosmological constant
\cite {Gogber,RandallSundrum}. 
The generalization of this framework
to higher dimensions 
\cite {ADDKaloper,CohenKaplan,LykkenRandall,Verlinde}, 
as well as a number 
of interesting  phenomenological and cosmological 
issues were studied in the literature 
\cite {general1,general2,general3,general4,Youm,Gubster}.  

As we mentioned above,
there is a localized graviton zero-mode
on the brane worldvolume \cite {RandallSundrum}.
In addition, there is a continuum of graviton states 
in the brane background \cite {RandallSundrum}. 
All the fields in 
anti de Sitter space\footnote{ 
Field theory in anti de Sitter space needs  careful 
treatment \cite {Avis}. In general, there are closed timelike 
curves in this space which can be  avoided by considering the 
universal cover  of ${\rm AdS}$ space \cite{Avis}. 
In the present work, we will adopt 
the simplified terminology calling the covering space
${\rm AdS}$.} should be given 
boundary conditions. 
This can be accomplished by adopting  the {\it holographic}
approach to the problem \cite {WittenSB}. In this framework, 
the bulk continuum is described by a certain Field Theory
on the brane\footnote{This, in fact, should be 
Conformal Field Theory  
since the isometry group of $\a$ (which is   
$SO(2,4)$) coincides with the conformal group of the 
brane (boundary) Minkowski space.}.
This field theory should be coupled to 
a four-dimensional graviton which is nothing but
the localized graviton zero-mode in the original $\a$ picture.
In a sense, the introduction of the brane  
generates  the coupling of the brane (boundary) four-dimensional 
field theory (which is expected  to describe 
the bulk gravitational physics in accordance with the Maldacena
conjecture 
\cite {Maldacena,WittenHol,PolyakovHol}),
to the four-dimensional Einstein gravity. 
This framework sets uniquely the 
boundary conditions for the fields (see 
detailed discussions and references in
\cite {WittenHol}). 
 
The aim of the present paper is to study localization of spin-
$0,1,1/2$ and $3/2$ fields on a 3-brane in the $\a$ space.
The study of these issues is crucial  for phenomenological
model-building as well as  
for the holographic description 
of the model. Besides, this can be useful for 
the task of realizing this scenario in a  supergravity framework. 

Another, seemingly different, but in fact closely related 
subject which  we will discuss here is the problem of the four-dimensional 
cosmological constant. 
We will argue that the necessity of the fine-tuning between the 
bulk cosmological constant and the brane tension can be avoided
by introducing a four-form gauge field.

In section 2 we study
localization of various spin fields  on a brane in $\a$ space.
We consider two different cases: a positive tension 
brane which localizes gravity \cite {RandallSundrum} (the 
Randall-Sundrum (RS) brane)
and a ``negative-tension'' brane which gives rise to
the exponentially growing warp factor\footnote{
The ``negative-tension'' brane is understood as a  
rigid non-dynamical slice of space with negative energy density.} 
\cite {Gogber}.
We show that a spin-$0$ massless field is localized on 
the RS  brane. The emerging picture is similar to 
that of the localized graviton. There is a scalar zero-mode
on the brane. In addition, there is a continuum of states 
with no mass gap. 
This fits well into the holographic approach mentioned  above.
Furthermore, we show that the ``negative tension brane'' with 
the exponentially rising warp factor does not allow to localize 
normalizable scalar zero-modes.

For spin-$1$ fields the conclusions are  less promising. Neither 
the positive-tension nor the ``negative-tension branes'' are capable of
localizing vector fields in the minimal setup. The way out for the 
phenomenological model-building would be to invoke
either stringy mechanism of localization on a D-brane worldvolume, 
or, in the field theory framework, 
to use the Dvali-Shifman
mechanism which is based on the bulk confinement \cite {DvaliShifman}. 
In either case  the introduction of new 
bulk physics is necessary. 
Furthermore, we show that in the minimal setup
neither spin-$1/2$ not spin-$3/2$ fermions are 
localized on the RS brane. 
Nevertheless, the localization of these fields can be 
achieved by introducing new interactions in the theory.
This allows one to have localized {\it chiral}
fermions on a brane.  
On the other hand, spin-$1/2$ and spin-$3/2$ fermions are localized
by gravitational interactions
on a  brane with ``negative tension'' which has the  exponentially 
rising warp factor.
A peculiar feature of this
mechanism is that the localized modes are not chiral.

In section 3 we discuss the fate of the four-dimensional cosmological 
constant. As we mentioned above, 
the scenarios of  Refs. \cite {Gogber,RandallSundrum} 
require a special fine-tuned relation of the bulk cosmological
constant to the brane tension. Once this 
relation is enforced, the four-dimensional brane 
Universe is static, i.e., the four-dimensional
cosmological constant is zero. Within the supergravity framework
this relation is just the BPS condition \cite {Cvetic}. 
However, one 
should not expect that the BPS  relation survives  
the quantum corrections after SUSY breaking.
As a result, one eventually goes back  to   
fine-tuning. 
We will show in section 3 
that the necessity of the fine-tuning can be removed from the 
theory. This is accomplished by introducing a four-form gauge 
field which couples 
to gravity only. 
The four-form field does not propagate 
physical degrees of freedom in five-dimensional space, 
however, it can give rise to
an arbitrary (negative or positive depending on the sign of 
its kinetic term) cosmological constant in the theory \cite 
{Duff1,Townsend}. 
Therefore, for any value of the original bulk cosmological constant,  
the net cosmological constant in this model  
is arbitrary. As a result, 
there are infinite number of brane Universe solutions. 
Each of these brane Universes 
is labeled by the corresponding 
four-dimensional cosmological constant.
Among all these possible brane Universes there is one 
with a zero cosmological constant. 
This removes the fine-tuning problem from the 
Randall-Sundrum scenario
in the  sense that a  static solution now exists
for any value  of the bulk cosmological constant and the 
brane tension.  
In addition, this framework  opens an opportunity 
for anthropic arguments.

Note that many properties of a 3-brane is $D=5$ are similar to those of
a 2-brane in $D=4$ which were previously studied  in \cite {Cvetic1}
(for a review see \cite {CveticRev}). 
These studies were recently generalized to $D-2$ branes (domain walls)  
in $D$ dimensions \cite {CveticW}.

\section{Localization via warp factors}

We start with the five-dimensional  gravity action and a 
cosmological constant (the $[+,-,-,-,-]$ signature 
will be assumed below) 
\beq
S=-M^3~\int d^5x \left ( R+2\Lambda \right )~.
\label{action1}
\eeq
In addition one includes a static 3-brane with tension $T$ 
which is located at $y=0$ \cite {RandallSundrum}. 
The five-dimensional interval will be parametrized  as follows: 
\beq
ds^2~=~A(y)~dx^2_{3+1}-dy^2~.
\label{metric}
\eeq
The Einstein equations in this case have 
two different solutions for the warp factor $A(y)$. 
For a positive-tension brane one finds \cite {RandallSundrum}
\beq
A(y)~=~ {\rm exp} \left (-H |y| \right )~,
\label{RS}
\eeq
while for a ``negative-tension'' brane the warp factor is exponentially 
rising\footnote{As before, the ``negative-tension'' brane is a  
non-fluctuating  slice of space with negative energy density. 
The exponentially rising  warp factor was originally 
introduced in Ref. \cite {Visser} in a theory with no branes.}
\cite {Gogber}
\beq
A(y)~=~{\rm exp} \left (H |y| \right )~.
\label{GV}
\eeq
In both cases, the static solutions (\ref {RS}) and (\ref {GV}) 
exist if  the bulk cosmological constant is  fine-tuned
to the brane tension
\beq
\sqrt {-{2\Lambda\over 3} }~=~{|T|\over 6M^3}~\equiv~H~.
\label{BPS}
\eeq
We will show below, that the RS  solution admits the 
localization of a  spin-$0$ state in addition to  the already known localization 
of gravitons.  The solution (\ref {GV}), on the other hand,  
localizes spin-$1/2$ and spin-$3/2$ fermions.

Before we go further, let us make some comments on the literature.
One can consider an  orbifold compactification
of the fifth dimension in the RS framework. In this case, 
there are two 3-branes which are located 
at the orbifold fixed points
\cite {RandallSundrum}. One of these is a positive-tension brane
and the other one is a negative-tension brane. 
The warp factor for one of them decreases as
in (\ref {RS}), while for the other one it increases exponentially
(\ref {GV}).  Therefore, in this framework 
spin-$0$ and spin-$2$ states 
will be localized on the  positive-tension brane, while 
spin-$1/2$ and spin-$3/2$ states will be trapped on 
the  negative-tension one.
In what follow we will be studying  
a single 3-brane which separates two patches of
$\a$ space-time. From the point of view  of the orbifold construction this
corresponds to the case when one of the orbifold fixed points 
is removed to infinity. 
  
\subsection{Spin-$0$ fields}

In this subsection  we study  the localization of a 
real scalar field in the background (\ref {metric}). We will 
find that the solution (\ref {RS}) admits a localized zero-mode.

Let us start with a massless scalar field  
coupled to gravity:
\beq
{1\over 2}\int d^5x \sqrt {g}g^{AB}\partial_A\Phi 
\partial_B \Phi~.
\label{scalar}
\eeq 
Here $A,B$ denote five-dimensional indices. 
The corresponding equation of motion takes the form
\beq
\partial_A\left (\sqrt{g}g^{AB} \partial_B\Phi \right) =0~.
\eeq
Using (\ref {metric}) and decomposing this equation in the 
four-dimensional and fifth dimensional parts one finds:
\beq
{1\over A}\eta^{\mu\nu}\partial_\mu\partial_\nu \Phi - 
{1\over A^2}\partial_y \left(A^2\partial_y\Phi \right) =0~.
\label{eom}
\eeq
Let us decompose the field $\Phi$ as follows:
\beq
\Phi(x,y)=\phi(x) ~\chi(y).
\eeq
A plane wave which propagates in a four-dimensional
worldvolume satisfies the corresponding four-dimensional 
equation of motion:
\beq
\eta^{\mu\nu}\partial_\mu\partial_\nu\phi(x)=-m^2\phi(x)~.
\eeq
Using this, and introducing a new variable 
\beq
u(y)=A(y)~\chi(y)~,
\label{u}
\eeq
one gets the following Schr\"odinger equation for the $y$-dependent
part:
\beq
\left [-\partial_y^2-m^2~e^{H|y|}+H^2-2H \delta(y) \right]~u(y)=0~.
\label{schrod}
\eeq
This equation coincides with the one for a localized graviton
\cite {RandallSundrum}. 
The zero-mass solution  ($m^2=0$)
to this equation takes the form:
\beq
u(y)=c~e^{-H|y|}~,
\eeq
where $c$ is a constant. 
It is interesting that, written in terms of the  original variable
$\chi$, the scalar zero-mode is just a constant
\beq
\chi (y) =c~.
\label{constant}
\eeq
At a first glance, such a solution could not  be localized since 
it is not suppressed  away from the brane. Nevertheless, 
the solution can still be considered as a localized mode.  
The presence of the exponential warp factor in 
the metric (\ref {metric}) allows one to perform the following 
decomposition for the zero-mode
\begin{eqnarray}
{1\over 2}\int d^4xdy \sqrt {g}~ g^{AB}~\partial_A\Phi_0 ~
\partial_B \Phi_0~=~{1\over 2}\int_{-\infty}^{+\infty}dy \sqrt {g} 
\chi^2(y)g^{\mu\nu}\int d^4x \partial_\mu\phi(x)
\partial_\nu\phi(x) \nonumber \\
 = {1\over 2}\int d^4x~\eta^{\mu\nu}\partial_\mu\phi(x) 
\partial_\nu\phi(x)~,
\label{scalarloc}
\end{eqnarray}
where we set $c=\sqrt{H/2}$. 

Thus, for the zero-mode $\Phi_0(x,y)=\sqrt{H/2}~\phi(x)$. Moreover,  
the field 
$\phi(x)$ is effectively ``localized'' on a  brane worldvolume
due to the exponentially decreasing warp factor.
Notice  that this zero-mode 
resembles a zero-mode of  the standard Kaluza-Klein compactification.
This is not  surprising since 
the effective size  of the fifth dimension, $L$, 
is finite, $L\sim 1/H$, 
in spite of the fact that this dimension  is not compact
\cite {ADDKaloper}.
Let us stress also  that such 
a constant localized solution is possible only for the  
warp factor (\ref{RS}), but not for (\ref{GV}). 

As in the case of gravitons \cite{RandallSundrum}, 
the equation (\ref {schrod}) gives rise to a tower
of continuum states with $m^2>0$. These states 
produce  a nonzero contribution of order $1/r^3$ to 
the usual four-dimensional $1/r$ potential mediated by a massless 
scalar exchange on a brane. The picture is similar to that of the 
localized graviton \cite {RandallSundrum}.
 
\subsection{Spin-$1$ fields}

There are no localized solutions in this case. 
Although there is a  constant solution similar to (\ref{constant}), 
nevertheless, the Lagrangian for a vector field 
does not yield  the  suppressing warp factor 
which was so crucial in the case of scalars.
Let us see this in some details. 
Using the definition of the 
5D field-strength
\beq
F_{AB}=\partial_A V_B-\partial_B V_A~,
\eeq
choosing the gauge $V_5=0$, and decomposing the remaining part 
of the massless vector field as 
\beq
V_\mu(x,y)=v_\mu(x)\sigma(y)~,
\eeq
we get ($f_{\mu\nu}\equiv \partial_\mu v_\nu-\partial_\nu v_\mu$)
\begin{eqnarray}
-{1\over 4}\int d^4xdy \sqrt {g}~g^{AB}~g^{CD}~F_{AC}~F_{BD}~
=~-{1\over 4}\int_{-\infty}^{+\infty}dy \sqrt {g} 
\sigma^2(y)g^{\mu\nu}g^{\alpha\beta}\int d^4x 
f_{\mu\alpha}f_{\nu\beta} \nonumber  \\
 = -{1\over 4}\int d^4x~\eta^{\mu\nu}\eta^{\alpha\beta}
f_{\mu\alpha}f_{\nu\beta}\int_{-\infty}^{+\infty}dy~\sigma^2(y)~.
\label{vectorloc}
\end{eqnarray}
This diverges  if $\sigma (y)$ is a constant. 

How about nontrivial solutions to the equation of motion 
for the vector field, could  they give localized  solutions?
One can check that such  normalizable solutions do not exist. 
Since this issue has already been studied in the orbifold version of
the RS scenario in \cite{Rizzo,Pomarol} we skip these derivations.
The net result is that neither (\ref {RS}) nor (\ref {GV})
localizes  massless vector fields.  

\subsection{Spin-$1/2$ fermions}

We start with the Lagrangian for massless spin-$1/2$ fermions 
\beq
i\sqrt{g}~{\bar \Psi} \Gamma^B D_B \Psi~.
\label{spin}
\eeq
The corresponding equation of motion can be decomposed as
follows:
\beq
\left (\Gamma^{\mu}~D_\mu ~+~\Gamma^5~D_5\right ) \Psi(x,y)=0~.
\eeq
The relations between the curved-space gamma matrices 
($\{\Gamma^A,\Gamma^B\}=2g^{AB}$) and the minkowskian ones 
($\{\gamma^A,\gamma^B\}=2\eta^{AB}$) read as follows:
\beq
\Gamma^{\mu}={1\over\sqrt{A}}\gamma^\mu~~,~~\Gamma^5=\gamma^5~.
\eeq
The spin-connection and 
covariant derivative can also be calculated for the metric 
(\ref {metric}): 
\beq
D_\mu~=~\partial_\mu+{A^{\prime}\over 4 A}~
\Gamma_\mu\Gamma^5~~,~~
D_5=\partial_5~.
\label{covariant}
\eeq 
After these conventions are set we can decompose the five-dimensional 
spinor into the four-dimensional and the fifth-dimensional parts:
$\Psi(x,y) =\psi(x) \alpha (y)$. We require that the four-dimensional part
satisfies the massless equation of motion $\gamma^\mu\partial_\mu\psi(x)
=0$.
As a result, we obtain  the following equation for the $y$ dependent part
\beq
\left ( \partial_y~+~{A^{\prime} \over A} \right ) \alpha(y) =0~.
\label{eqnferm}
\eeq
The solution to this equation reads:
\beq
\alpha(y)={c\over A(y)}~.
\eeq
This  is not normalizable in the RS case (\ref{RS}), but it is 
normalizable if the warp factor (\ref{GV}) is used. 
To see this, one has to decompose  the action (\ref{spin})
as follows:
\begin{eqnarray}
\int d^4x\int_{-\infty}^{+\infty} dy\sqrt{g}{\bar\Psi}(x,y)
i\Gamma^BD_B\Psi (x,y)= \nonumber\\
\int_{-\infty}^{+\infty}dy A^{3/2}(y)\alpha^2(y)
\int d^4x{\bar\psi}(x)i\gamma^\mu\partial_\mu\psi(x)~,
\nonumber
\end{eqnarray}
which is infinite for $A(y)$ defined in  (\ref{RS}),  but is finite for 
$A(y)$ from (\ref{GV}) (we choose $c=\sqrt{H}/2$ to get the correct 
normalization)\footnote{After we obtained these results 
the work \cite {neubert} appeared, in which  it was shown that 
in the orbifold RS construction  massless spin- 
$1/2$ fields are not localized on a brane.}.
 
In order to localize spin-$1/2$ fermions in 
the RS framework  one could  use the method 
of localization due to Jackiw and Rebbi \cite {JackiwRebbi}.
For this, one introduces the interaction  of fermions with 
a scalar field, $\Phi {\bar\Psi} \Psi$. 
The scalar should interpolate between two different vacua 
at different sides of the brane (one could take the kink solution, 
for instance). This gives  an effective  five-dimensional
mass to fermions (constant fermion mass approximation is good enough
when the scalar is heavy). However, the 
five-dimensional fermion mass term $M {\bar \Psi}\Psi$
would flip  the sign under the reflection with respect to the brane. 
Under these circumstances, the chiral left(right)-handed component
of the fermion can be localized on the brane \cite {JackiwRebbi}. 
The equation of motion for the fermion takes the form:
\beq
\left ( \partial_y+{A^{\prime} \over A} 
+i\gamma_5 M~[\theta(y) -\theta(-y)]~
\right ) \Psi(x,y) =0~,
\label{eqnferm1}
\eeq
where $\theta(y)$ is the step-function. 
Introducing the chiral modes as 
$i\gamma_5\Psi_{L,R}=\pm \Psi_{L,R}$,
one finds that the chiral 
solution $\Psi_L \propto {\rm exp} (H-M)$ is localized
for (\ref {RS}) as long as:
\beq
M~>~ {H\over 4}~.
\eeq 
Likewise, in the background (\ref {GV}), the mode $\Psi_L$
is  localized if $M<H/4$. 

\subsection {Spin-$3/2$ fermions}

The consideration for gravitinos is similar to that of 
spin-$1/2$ fermions. Therefore, 
our discussions will be brief.
The equation of motion for a massless gravitino reads as:
\beq
\Gamma^{[A}\Gamma^{B}\Gamma^{C]}~D_B~\Psi_C=0~.
\eeq 
Here, the square brackets denote antisymmetrization w.r.t.
all indices. We choose the gauge $\Psi_5=0$ and split 
the remaining fields  as  $\Psi_\mu(x,y) =\psi_\mu (x)~u(y)$. 
Using now the four-dimensional 
gauge choice $\gamma^\mu\psi_\mu=\partial^\mu\psi_\mu=0$,
and the four-dimensional  equation of motion for a  massless 
spin-$3/2$ field 
$\gamma^{[\mu}\gamma^\nu\gamma^{\sigma]}\partial_\nu\psi_\sigma=0$, 
one finds the following equation for the $y$ dependent part:
\beq
\left ( \partial_y+{A^{\prime} \over 2A} \right ) u(y) =0~.
\label{eqnferm2}
\eeq
The solution to  this equation is
\beq
u(y) ={c\over \sqrt{A(y)}}~.
\label{solgrav}
\eeq
In the RS case this is  not a normalizable 
function. The action 
\begin{eqnarray}
\int d^4x\int_{-\infty}^{+\infty} dy\sqrt{g}{\bar\Psi}_A(x,y)
i\Gamma^{[A}\Gamma^B\Gamma^{C]}D_B\Psi_C(x,y)= \nonumber \\
\int_{-\infty}^{+\infty}dy A^{1/2}(y)u^2(y)
\int d^4x{\bar\psi}_\mu(x)i\gamma^{[\mu}\gamma^\nu\gamma^{\sigma]}
\partial_\nu\psi_\sigma(x)~, 
\label{gravotinoloc}
\end{eqnarray}
diverges because of  the $y$ integration. On the other hand, 
for the exponentially rising warp factor (\ref {GV}), 
the $y$ integral is finite and the 4D action is 
canonically  normalized for $c=\sqrt{H}/2$. Therefore, the solution (\ref{GV}) 
admits a localized free spin-$3/2$ zero-mode on the 3-brane.
Notice that this mode is not chiral.

Finally, let us comment on a possibility   
of localization of gravitinos 
on the background (\ref {RS}).
One should introduce some additional 
interactions for this to happen (as we did for the case of 
spin-$1/2$ fields). The simplest way would be 
to realize the RS solution (\ref {RS}) in 
some $N=2$ five-dimensional supergravity.
In this case, one could hope to find a BPS 3-brane which preserves
half of the original supersymmetries. 
Since the background (\ref {RS}) localizes gravitons, by supersymmetry,
its SUGRA counterpart would  also localize gravitinos.
The dynamical reason for the localization of gravitinos
could  be their interactions with some other fields of 
the corresponding SUGRA. 
Detailed studies  of these issues 
could be a subject of a separate project.

\section{Cosmological Constant on a Brane}

In this section we study what happens if the bulk cosmological constant
is {\it not} fine-tuned to its critical value which is defined by  Eq.
(\ref {BPS})\footnote{We are grateful to Gia Dvali for 
many useful discussions of the results of this section.}. 
This is expected to be the generic case in any  realistic 
brane Universe models with broken SUSY. 
Indeed, after  SUSY is broken, 
equation (\ref {BPS}) can  no longer be protected against corrections 
even though it could have been obtained as a BPS equation 
in some five-dimensional theory of supergravity ( 
along the lines of  \cite {Cvetic}, for instance).   
Therefore, in what follows we consider 
the case when the bulk cosmological constant differs  from its
critical value defined by  (\ref {BPS}).
In such a case  the four-dimensional brane Universe is not static
any more \cite {N,Kaloper}. The four-dimensional cosmological constant
is determined  by the difference between the actual bulk cosmological constant
and the critical cosmological constant satisfying (\ref {BPS}).
Generically, this difference can be big, leading to 
a de Sitter or anti de Sitter   
four-dimensional brane Universe with an unacceptably large 
four-dimensional cosmological constant.
Thus, it seems that the fine-tuning is of vital importance. 
However, the fine-tuning of the parameters can be avoided 
by introducing into the theory a four-form gauge field
$A_{BCDE}$.  
This field cannot propagate any physical degrees of freedom 
in five-dimensions.
Nevertheless, it gives rise  to a  dynamically generated 
cosmological constant \cite {Duff1,Townsend}.
The value of this cosmological constant is arbitrary, since it appears 
as a constant of integration of  the equation of motion for 
the $A_{BCDE}$ field.
Therefore, the net cosmological constant in five-dimensions
is the sum of the original cosmological constant 
and the dynamically generated one\footnote
{Notice that it is important in our case that the four-form 
field does not couple to 
the brane itself, i.e., the brane does not carry the 
corresponding ``Ramond-Ramond charge'', see discussions in the next section.}. 
Let us for simplicity assume that the original bulk cosmological constant
$\Lambda$ is positive but otherwise an arbitrary 
number which is not necessarily fine-tuned to the brane tension
(the generalization for 
negative  $\Lambda$ is straightforward and will be given  in the next section).
The action for the system can be written as follows: 
\beq
S=-M^3 \int d^5x \sqrt{ g}~\left (R~+~2\Lambda \right )+ \int d^5x \sqrt{ g} 
\left \{-{1\over 2 \times 5!}
F_{ABCDE}F^{ABCDE} \right \} +S_{\rm Brane}~.
\label{action}
\eeq
Here $F_{ABCDE}$ denotes the gauge-invariant field-strength 
for the field $A_{BCDE}$ (the choice of the  sign of the 
kinetic term for this field will be discussed in the 
next section).
The action for the brane itself in 
(\ref {action}) is denoted by $S_{\rm Brane}$.
We will simply assume that the 3-brane is a static source
which is localized at $y=0$.  It  has the following
energy-momentum tensor:
\beq
T^{\rm Brane}_{AB}=T~ \delta(y) ~{\rm diag} (1,-1,-1,-1, ~0)~, 
\label{tensorB}
\eeq
where $T$ denotes the tension of the 3-brane. 
  
The equations of motion of the system take the following form:
\begin{eqnarray}
R_{AB}-{1\over 2} g_{AB} R ={1\over 2 M^3}~ \left 
(~T_{AB}~+~T^{\rm Brane}_{AB}~\right )~+~g_{AB}~\Lambda~,
 \nonumber \\
\partial_A \left (\sqrt {g}~F^{ABCDE}(y)\right )~=~0~.
\label{eqmo}
\end{eqnarray}
Here $T_{AB}$ is the energy-momentum tensor
for the four-form field:
\beq
T_{AB} \equiv {1\over 4!}~ \left \{ 
-F_{ACDEG}F_{B}^{~CDEG}+{1\over 10} ~g_{AB}~F_{CDEGH}F^{CDEGH}  \right \}~. 
\label{tensorA}
\eeq
As we discussed above, 
the crucial property of  the four-form gauge field in five-dimensional 
space-time  is that it does not propagate dynamical degrees of freedom.
The whole dynamics is  eliminated
by the system of equations of motion,  and the gauge constraints which  
emerge as a result of the gauge invariance of the action. 
Nevertheless, this system of equations has 
a constant field-strength solution
which gives rise to 
a  nonzero vacuum energy density. 
This is what happens in general with a d-form 
gauge invariant field-strength in d-dimensional space-time. 

To make these discussions quantitative let us 
look for the 
following solution to  the equations of motion (\ref {eqmo}):
\begin{eqnarray}
ds^2=~{\cal A}(y)\left (~ dt^2~-~a(t)~dx^idx^i ~\right )~-~dy^2~,
\label{interval}
\end{eqnarray}
where $a(t)$ is 
the scale factor of the four-dimensional brane Universe.
The corresponding background for the four-form field is given by
\beq
F^{ABCDE}={1\over \sqrt {g}} ~\varepsilon^{ABCDE}~k~,
\label{ansatz} 
\eeq
where  $k$ stands for an arbitrary integration 
constant (in our normalizations $\varepsilon^{12345}=1$). 
With this {\it Ansatz} at hand  the four-dimensional components 
of the energy-momentum tensor for the 
four-form field take  the form:
\beq
T_{AB} = -{1\over 2} ~g_{AB}~ k ^2~.
\label{rr}
\eeq
This generates an additional negative contribution to the 
total  effective  bulk cosmological constant 
\beq
\Lambda_{\rm eff}~ =~\Lambda~-~{1\over 4 M^3}~ k^2~.
\label{cc} 
\eeq
Using these expressions the Einstein equations can be written as
follows:
\begin{eqnarray}
{3 {{\cal A}^{\prime\prime}} \over 2 {\cal A}}~=
-~\Lambda_{\rm eff}-{1\over 2 M^3} T 
\delta (y)~,~~~~~~~~~~~~~~~
\Big ( { {{\cal A}^{\prime} } \over {\cal A}}\Big )^2 ~=~{H_0^2\over 
{\cal A}^2}~-{2\over 3} \Lambda_{\rm eff}~,\nonumber \\
\left ( { {\dot a}(t)\over a(t)}  \right )^2= {{\ddot a}(t) \over a(t)}~
\equiv~H_0^2~.~~~~~~~~~~~~~~~~~~~~~~~~~~~~~
\label{einstein1}
\end{eqnarray}
Here, primes  denote differentiation with respect to $y$,
and dots denote differentiation w.r.t. the time variable.
The solution to these equations reads as \cite {N,Kaloper}\footnote{
For simplicity we present here the solution 
when the brane worldvolume is ${\rm dS}_4$. The solution can be obtained 
for ${\rm AdS}_4$ as well \cite {N,Kaloper}.}
\begin{eqnarray}
{\cal A}(y)~=~{\rm ch}\left (\sqrt { -{2\Lambda_{\rm eff}\over 3}} y  \right) 
-{T \over 2M^3 \sqrt { -{6\Lambda_{\rm eff}}} }~{\rm sh}
\left (\sqrt { -{2\Lambda_{\rm eff}\over 3}} |y|  \right)~.
\label{sol1}
\end{eqnarray}
The four-dimensional 
Hubble constant is defined as follows:
\beq
H_0=\sqrt{ {T^2\over 36M^6}+{2 \Lambda_{\rm eff} \over 3} }~. 
\label{hubble}
\eeq
Therefore, the four-dimensional 
cosmological constant, $\Lambda_4\equiv H_0^2$,   
contains an  arbitrary integration constant $k$
\beq
\Lambda_4~ =~{T^2\over 36M^6}+{2 \over 3}~ \left 
( ~\Lambda~-~{1\over 4 M^3}~ k^2~   \right )~.
\label{k}
\eeq
Let us summarize the results of our discussions. 
The system (\ref {action}) {\it without} the $F$ field
does not allow for a static solution 
for a generic, non-fine-tuned values of the bulk cosmological 
constant and the brane tension.  
However, when  the  $F$ field is switched on, 
there  are  an infinite number
of brane Universe solutions. These Universes differ 
from each other by the value of the 
corresponding four-dimensional cosmological constants
defined in (\ref {k}). Among these solutions there is the 
static brane with zero four-dimensional cosmological constant. 
This removes the necessity of the fine-tuning  
in the sense that static solution now exists for any 
values of $\Lambda$ and $T$.
Certainly, the cosmological constant problem is not 
solved, instead, 
the introduction of the $F$ field
gives an opportunity for the anthropic arguments.
Indeed, only those branes 
for which the four-dimensional cosmological 
constant fits into the Weinberg's  window \cite {Weinberg}:
\beq
{\Lambda_4 \over 8\pi G_N}~ \leq ~\left (~ 10^{-3}~{\rm eV} 
~\right )^4~,
\label{window}
\eeq
could  accommodate  the Universe similar to ours.
All other brane Universes, not satisfying (\ref {window}), 
would not be able to form the large scale structures.
Concluding, the action (\ref {action}) allows to avoid 
the necessity of fine-tuning of the bulk cosmological constant $\Lambda$ to 
the brane tension $T$,  for any
$\Lambda$ and $T$ there are  
an infinite number of possible brane Universe solutions 
of (\ref {action}) and one of these solutions has zero four-dimensional 
cosmological constant.

\section{Discussions and conclusions}

Let us start with  some comments on the form of 
the action (\ref {action}). The four-form field in (\ref {action})
reminds the Ramond-Ramond (RR) field to which a three-brane could  couple.
However, one can check that the sign of the kinetic term in 
(\ref {action}) is inverse to what should have been used for the
RR  field. The choice of this sign is not restricted by 
positivity arguments in the five-dimensional theory (\ref {action}),  
since this field 
does not propagate dynamical degrees of freedom. 
However, the negative sign is necessary in (\ref {action})
if the four-form field is {\it real} and one needs to generate   
a {\it negative} cosmological constant in (\ref {cc}).
This is true as long as the original cosmological constant
$\Lambda$  in (\ref {action}) is an arbitrary positive number. 
Thus, for $\Lambda>0$ the four-form field cannot be thought of 
as a RR field. Let as now see what happens if $\Lambda$
is a big  arbitrary negative number (in fact, without
loss of generality, we take its 
magnitude to be bigger than the magnitude of the critical 
value defined by (\ref {BPS})).
In this case one can just 
flip the sign of the kinetic term 
of the $F^2$ term in (\ref {action}) and obtain the desired result\footnote{
Alternatively, one could  add to the action (\ref {action}) 
another four-form field with opposite kinetic term.}.
Indeed, in this case, the kinetic term for the four-form field is precisely 
equivalent to that for a  RR field. 
This produces a positive cosmological constant
in the bulk. Therefore, the expression (\ref {k}) takes the form:
\beq
\Lambda_4~ =~{T^2\over 36M^6}+{2 \over 3}~ \left 
( ~\Lambda~+~{1\over 4 M^3}~ k^2~   \right )~.
\label{nk}
\eeq
Since $\Lambda$ in this case is  a big negative number
(its magnitude is bigger than $T^2/24M^6$), 
one is still able to 
have an infinite number of  brane Universe solutions
parametrized by the integration constant $k$
and one of these solutions have $\Lambda_4=0$.
In this case (negative $\Lambda$),
one could think of the four-form field   
as some kind of RR field of string theory. However, in our case 
the brane at hand should not couple to this RR field, i.e.,
the brane should not be a D-brane. Otherwise, the effective 
cosmological constant produced by the RR field would be defined by the 
brane RR charge. Since this is quantized, one would again need fine-tuning
between the brane charge and the brane tension (as before, in the supergravity 
framework this is the BPS condition, however it  becomes 
fine-tuning after  SUSY is broken). 

Let us now go back to the case $\Lambda>0$ considered in 
the previous section. 
A way to think of the four-form field  in this case 
is to recall that in theories of supergravity
there are auxiliary scalar fields. 
These scalars  usually enter an off-shell Lagrangian 
in a quadratic form without kinetic terms. 
One of these scalars, let us call 
it $\phi$, can be dualized into the five-form field strength via
$\phi\propto \varepsilon_{ABCDE} F^{ABCDE}$. Since  the original 
scalar was an auxiliary field and the four-form does not propagate 
dynamical degrees 
of freedom either, this dualization makes sense. As a result,
the $\phi^2$ term in the original off-shell Lagrangian of supergravity,
which can emerge as $\sqrt{g}(R+\phi^2)$ \cite {zuker},  
is  replaced by the $F^2$ term  in the form given  in (\ref {action}).
This is acceptable as far as bosonic modes are concerned. 
The actual issue is whether it is possible to perform  the dualization of the 
whole SUGRA supermultiplet, not only its scalar part. 
Such a dualization of the chiral $N=1$ SUGRA supermultiplet 
into the tensor supermultiplet  
was shown to be possible in the four-dimensional case
\cite {Waldram}. The issue whether
the same  can be done in five-dimensions remains open. 

To conclude, we studied localization of various spin fields 
on a brane in $\a$. We found  that spin-$0$ 
fields are localized on a  brane with  positive tension
which also localizes  gravitons.
The spectrum of spin-$0$ field is similar to that of the localized 
graviton: there is a single
localized zero-mode plus  a continuous tower 
of states. This continuum can be thought of as a bulk five-dimensional field.
To fix the boundary conditions, 
the whole picture should be viewed in a  sense of the 
AdS/CFT correspondence.
Spin-$1$ fields are not localized neither on a  
brane with  positive tension
nor on a   brane with  
``negative tension''. Within the field theory framework  
the Dvali-Shifman mechanism \cite {DvaliShifman} 
should be invoked for the vector field  localization.
In string theory, vector fields could  be localized by 
assuming that the brane at hand is in fact a D-brane.
Spin-$1/2$ and spin-$3/2$ fields are localized due to 
gravitational interactions on a   brane with  
``negative tension''. The localized modes in this case are not chiral. 
In order to trap
spin-$1/2$ and spin-$3/2$ chiral zero-modes  
on a  positive-tension brane 
additional fields should be introduced.  
These could  be scalars of five-dimensional gauged supergravity.

Finally, we have shown that  fine-tuning between the 
bulk cosmological constant and the brane tension can be avoided  
if there is a four-form gauge field in the theory. This field 
gives rise to an additional contribution to the bulk cosmological constant.
As a result, the model  with any generic $\Lambda$ and $T$ 
supports  an infinite number  of brane Universe
solutions which are parametrized by the continuous family of four-dimensional 
cosmological constants. Among these solutions there is  one 
with $\Lambda_4~=~0$.   

\vspace{0.2in}

The authors are grateful to Gia Dvali and Massimo Porrati 
for many useful discussions. 
We thank Goran Senjanovi\'c for reading the manuscript
and useful suggestions.
We are grateful to  Antonio Grassi, Juan Maldacena, 
Giuseppe Policastro and 
Hossein Sarmadi for helpful conversations.
The work was supported by the grant NSF PHY-94-23002.
The work of BB was also supported by the Ministry of Science and Technology
of the Republic of Slovenia.

\end{document}